\documentclass[aps,pra,reprint,floatfix]{revtex4-2}

\usepackage{amsmath}
\usepackage{amsfonts}
\usepackage{amssymb}
\usepackage{physics}
\usepackage{graphicx}

\newcommand{\SA}{_\textsc{a}}
\newcommand{\SB}{_\textsc{b}}
\newcommand{\SAB}{_\textsc{ab}}
\newcommand{\SV}{_\textsc{v}}

\begin{document}

   \title{Heat pump driven entirely by quantum correlation}
   \author{Tharon Holdsworth}
   \author{Ryoichi Kawai}
   \affiliation{Department of Physics, University of Alabama at Birmingham, Birmingham, Alabama 35294, USA}

   \date{\today}

   \begin{abstract}
   The second law of thermodynamics prohibits spontaneous heat from a cold to a hot body. However, it has been theoretically and experimentally shown that energy can flow from a cold to a hot body if the bodies are initially correlated. We investigated the \emph{anomalous energy exchange} between dissipation-less quantum systems that are initially entangled.  Then, we extended this model to include dissipation demonstrating \emph{anomalous heat} from a cold to a hot body. Based on these models, we constructed a heat pump driven entirely by quantum correlation as fuel and investigated its performance with numerical simulations.  Using the recently proposed definition of efficiency based on mutual information, the performance of the pump is found to be consistent with the second law of thermodynamics.
   \end{abstract}

   \maketitle

   \section{Introduction}
   In 1854, Rudolf Clausius stated that ``Heat can never pass from a colder to a warmer body without some other change, connected therewith, occurring at the same time."\cite{Clausius1854}, laying the foundation for the second law of thermodynamics and endowing entropy with its monotonically increasing quantity. Modern interpretations of the Clausius statement tell us that when heat flows spontaneously from a warmer to a colder body, entropy irreversibly increases. If heat were to spontaneously flow in the opposite direction, entropy would have to decrease, contradicting the second law. This notion of irreversibility suggested by the second law extends beyond thermodynamics, as it is also possible to gauge the ``one directional'' nature of time on the basis of increasing entropy \cite{Halliwell1994,Zeh2001,Parrondo2009}.

   However, Clausius' statement does allow heat to flow from a colder  to a warmer body \emph{with some other change}. For example, a heat engine that converts some of the heat flowing from a warmer to a colder body into extractable work. Onsager's reciprocal theorem tells us that if the direction of time is reversed for the heat engine, work added to the engine will drive heat from the colder to the warmer body. In this example work from an external source is the \emph{some other change} in Clausius' statement that can reverse the natural direction of heat.

   Since the introduction of quantum information and resource theories, it has been shown that quantum coherence can be a source of thermodynamic work \cite{Vinjanampathy2016,Bera2017,Levy2018,Latune2021}. Various thermodynamic machines driven or enhanced by quantum coherence and quantum measurement have been proposed, that seemingly violate the thermodynamic second law \cite{Scully2003,Elouard2017,Yi2017,Brunelli2017,Elouard2018,Ding2018,Hewgill2018,Bresque2021,Campisi2017,Buffoni2019}. To account for the role of quantum coherency in the thermodynamics context, the field of quantum thermodynamics is being developed at the intersection of thermodynamics and quantum information theory \cite{Kosloff2013,Vinjanampathy2016,Binder2018}. Quantum coherence is becoming a key thermodynamic resource for quantum thermodynamic devices.

   In particular, it has been shown that a certain \emph{initial} correlation between two systems in independent thermal states can induce energy flow against the temperature gradient \cite{Partovi2008,Jennings2010,Jevtic2012,Levy2018,Pyharanta2022}.   The thermodynamics arrow of time and fluctuation theorems for initially correlated systems have also been discussed \cite{Partovi2008,Jennings2010,Micadei2020}.
   Recently, Micadei \textit{et al.} \cite{Micadei2019} experimentally demonstrated that energy can be spontaneously transferred from a cold to a hot nuclear spin if their initial states are locally in a thermal (Gibbs) state with different temperature but entangled. Since dissipation does not take place in the isolated spins, the energy exchange between them cannot be considered heat. However, their experimental data has clearly shown the anomalous energy exchange caused by the interplay of interaction Hamiltonian and initial quantum correlations.

   Inspired by the work of Micadei \textit{et al.}, we propose a heat pump driven purely by quantum coherence. No energy is added directly to the heat pump through external work nor quantum measurement; heat spontaneously flows from a cold to a hot heat bath. This does not imply the heat pump violates the second law of thermodynamics since the quantum entanglement must be injected to the pump each cycle as fuel. By taking into account mutual information as a part of thermodynamic entropy, we show that the heat pump still operates below the Carnot efficiency.

   This paper is organized as follows:  Section \ref{sec:thermal} briefly reviews anomalous heat introduced by the previous works \cite{Partovi2008,Jennings2010,Jevtic2012,Micadei2019}. Section \ref{sec:qubits} presents our analysis of anomalous energy exchange between a pair of qubits isolated from environments. In the following section,
   the model is extended to include infinitely large heat baths which tend to decohere the qubits coupled to them. In Sec. \ref{sec:heat-pump}, we propose a  heat pump based on the anomalous heat conduction discussed in the preceding sections and report on its simulated performance.  In the final section, we briefly discuss possible mechanisms of generating an entangled pair of qubits that are locally in thermal states necessary to fuel our heat pump.

   \section{Thermal processes starting from a state locally in equilibrium but globally correlated}\label{sec:thermal}

   When a system is initially  in thermal equilibrium, its time evolution is subject to various restrictions \cite{Partovi2008,Jennings2010,Jevtic2012}. For example, it is not possible to extract work from a system in thermal equilibrium by any periodic process, as the Kelvin-Plank statement indicates.  When two systems are independently in thermal equilibrium at differing temperatures, their time evolution is again restricted. As the thermodynamic second law states, heat must flow from the hot to the cold body when the two systems are coupled. However, if the systems are initially away from thermal equilibrium, there are no such restrictions. We consider a special situation where two isolated systems are individually in local thermal equilibrium but the composite state of the two systems is initially correlated and thus out of equilibrium. We will find some thermodynamic restrictions are relaxed in the presence of initial correlations that allow the direction of heat to be reversed.

   Consider an isolated system comprised of two subsystems, \texttt{A} and \texttt{B}, which can be infinitely large or as small as a pair of qubits.  The composite state of the total system is denoted as $\rho\SAB$ and the local states of each subsystem are given by the reduced density operators $\rho\SA = \Tr\SB \rho\SAB$ and $\rho\SB = \Tr\SA \rho\SAB$. The Hamiltonian of each subsystem is notated as $H\SA$ and $H\SB$ so the energy of each subsystem is then given by $E\SA = \Tr\SA (H\SA \rho\SA)$ and $E\SB = \Tr\SB (H\SB \rho\SB)$, respectively. Since the composite system is isolated, the total energy $E\SA+E\SB$ remains fixed. (For simplicity, the interaction energy between the subsystems is ignored in the spirit of the thermodynamics setting but will be considered in the proceeding sections.)

   Such an isolated system evolves under a unitary transformation that conserves the total von Neumann entropy $S\SAB = - \Tr( \rho\SAB \ln \rho\SAB)$ where the Boltzmann constant $k_\textsc{b}=1$ is assumed. On the other hand, the sum of  the subsystem entropies, $S\SA = - \Tr\left(\rho\SA \ln \rho\SA\right)$ and $S\SB = - \Tr\left(\rho\SB \ln \rho\SB\right)$, is not conserved in the presence of correlation between them.  We stress that these entropies do not necessarily represent thermodynamic entropy, rather, each subsystem entropy should be considered as a measure of information content at a given time. The correlation between the subsystems can be quantified by the mutual information $I\SAB = S\SA  + S\SB - S\SAB$, also equivalently defined by the relative entropy $I\SAB = S(\rho\SAB\|\rho\SA\otimes\rho\SB)$ where $S(\rho\|\sigma) = \Tr(\rho\ln\rho)-\Tr(\rho\ln\sigma)$. By Klein's inequality \cite{Nielsen2010}, the relative entropy is strictly non-negative and vanishes if and only if $\rho=\sigma$, implying that the mutual information is always non-negative and vanishes only if the composite state is in a separable state $\rho\SA\otimes\rho\SB$.

   Let us assume that subsystems \texttt{A} and \texttt{B} are initially in thermal equilibrium at inverse temperature $\beta\SA = 1/T\SA$ and $\beta\SB = 1/T\SB$, respectively. The corresponding local densities are $\rho\SA^0 = e^{-\beta\SA H\SA}/Z\SA$ and $\rho\SB^0 = e^{-\beta\SB H\SB}/Z\SB$, where $Z\SA$ and $Z\SB$ are the respective subsystem partition functions. Inverting the Gibbs state as $H\SA = - T\SA(\ln \rho\SA^0 + \ln Z\SA)$, the energy of subsystem \texttt{A} can be expressed as $E\SA(t) = -T \SA \Tr\SA\{\rho\SA(t) \ln \rho\SA^0\} - T\SA \ln Z\SA$ and similarly for subsystem \texttt{B}. The deviation of the energy and entropy from their initial values must satisfy
   \begin{equation}\label{eq:dE-dS}
      \beta\SA\Delta E\SA (t) - \Delta S\SA = S(\rho\SA(t)\|\rho\SA^0) \ge 0,
   \end{equation}
   where $\Delta E\SA(t) \equiv E\SA(t) - E\SA(0)$ and $\Delta S\SA(t) \equiv S\SA(t) - S\SA(0)$ likewise subsystem \texttt{B} takes a corresponding expression.

   Although Eq. \eqref{eq:dE-dS} resembles the second law of thermodynamics, it does not require a thermal environment and is valid for any subsystem initially in a Gibbs state. Using the conservation of total energy, $\Delta E\SA(t) + \Delta E\SB(t)=0$, and total entropy, $\Delta S\SA(t)+\Delta S\SB(t)=\Delta I\SAB(t)$, Eq. \eqref{eq:dE-dS} leads to
   \begin{equation}\label{eq:dEA}
     \left(\beta\SA-\beta\SB\right) \Delta E\SA(t) =S(\rho\SA(t)\|\rho\SA^0) + S(\rho\SB(t)\|\rho\SB^0) + \Delta I\SAB(t),
   \end{equation}
   where $\Delta I\SAB = I\SAB(t)-I\SAB(0)$.

   Let us now assume that $T\SA > T\SB$. Equation \eqref{eq:dEA} indicates that if the subsystems are always uncorrelated [$\Delta I\SAB(t)=0$], $\Delta E\SA$ is negative at all times and thus energy is always transferred from the hot to the cold subsystem (we shall call this  \emph{normal} energy exchange), as expected from the standard theory of thermodynamics. Similarly, if $\Delta I\SAB(t) \ge 0$ the energy flows in the normal direction but the amount of transferred energy is larger than the normal energy exchange.  On the other hand, if the mutual information decreases significantly, that is $\Delta I\SAB(t) < -\left[S(\rho\SA(t)\|\rho\SA^0) + S(\rho\SB(t)\|\rho\SB^0)\right] < 0$, then energy flows from the  cold to the hot subsystem against the temperature gradient, (we shall call this \emph{anomalous} energy exchange), which is unexpected in the standard theory of thermodynamics.

   It is misleading to say that the above argument represents a violation of the second law. Temperature is simply a parameter specifying the initial energy distribution and  Eq. \eqref{eq:dEA} is valid for any type of evolution, unitary or nonunitary. It merely shows that at a certain time the hot subsystem \texttt{A} can reach a higher-energy state than its initial state but does not have to remain in the higher-energy state, in fact the energy can oscillate in time and $\Delta E\SA(t)$ can be positive or negative at different points in the evolution if the system is finite.  Strictly speaking, only when the subsystems are infinitely large, can $\Delta E\SA$ and $\Delta E\SB$ dissipate as heat $Q\SA$ and $Q\SB$.

   Regardless of system size, the above discussion has significant ramifications, namely, that the energy of a hot body can get even higher than its initial energy for a time if and only if the system is initially correlated. In light of this conclusion, we show that particular initial conditions can reverse the direction of energy flow against the temperature gradient or enhance the energy flow in the natural direction for a brief period of time before the initial correlations are lost to decoherence. In Sec. \ref{sec:heat-pump}, we exploit this effect to construct a heat pump driven purely by quantum entanglement.

   To illustrate the mechanical origin of the anomalous energy exchange, assume that the initial state of the two qubits \texttt{A} and \texttt{B} is given by
   \begin{equation}\label{eq:initial-rho}
      \rho\SAB^0 = \rho\SA^0 \otimes \rho\SB^0 + \chi,
   \end{equation}
   where $\chi$ is the correlation matrix describing all classical and quantum correlations.

   When the two subsystems are coupled through an interaction potential $V\SAB$, energy is exchanged between the subsystems. We are particularly interested in the time evolution of the subspace energy difference (SED) $\Delta\SAB(t) = E\SA(t) - E\SB(t)$.
   By expanding it in the Taylor series up to the first order of $t$, we find its initial trend:
   \begin{equation}\label{eq:DE-slope}
      \Delta\SAB(t) = \Delta\SAB(0) - i \Tr\left\{(H\SA-H\SB)\comm{V\SAB}{\chi}\right\} t + O(t^2),
   \end{equation}
   where $\comm{\cdot}{\cdot}$ is the commutator. If the first-order term is positive, the SED increases, indicating that the hot subsystem gets even hotter and the cold subsystem gets even colder. Thus, $\comm{V\SAB}{\chi}\ne 0$ is a necessary condition for the anomalous energy exchange.
   While the initial trend \eqref{eq:DE-slope} can be nonzero for any type of correlation satisfying the condition, we focus on quantum entanglement as a source of correlation.

   \section{Anomalous energy exchange between qubits}\label{sec:qubits}

   To explicitly demonstrate the anomalous energy exchange, we consider a pair of identical qubits with Hamiltonians $H\SA=H\SB=\frac{\hbar\omega}{2}\sigma_z$.  The ground and excited states are denoted as $\ket{0}$ and $\ket{1}$, respectively. For simplicity, we assume that $\hbar \omega = 1$ and all energies are normalized by $\hbar \omega$. Initially the qubits are disconnected and in independent Gibbs states with respective temperatures $T\SA > T\SB$. The uncorrelated part of density matrix in Eq. \eqref{eq:initial-rho} can be expressed in the product basis of two qubits $\{\ket{00}, \ket{01}, \ket{10}, \ket{11}\}$:
   \begin{equation}\label{eq:product-state}
   \rho\SA^0 \otimes \rho\SB^0=
   \begin{pmatrix}
      \lambda_1 &         0 &    0&       0\\
      0 & \lambda_2 &      0 & 0\\
      0 &         0 & \lambda_3 &      0\\
      0 &         0 &         0 & \lambda_4
   \end{pmatrix},
   \end{equation}
   where the matrix elements are defined by
   \begin{eqnarray}
      \lambda_1 &=& e^{(\beta\SA+\beta\SB)/2}/(Z\SA Z\SB), \nonumber\\
      \lambda_2 &=& e^{(\beta\SA-\beta\SB)/2}(Z\SA Z\SB), \nonumber\\
      \lambda_3 &=& e^{(-\beta\SA+\beta\SB)/2}/(Z\SA Z\SB), \nonumber\\
      \lambda_4 &=& e^{-(\beta\SA+\beta\SB)/2}/(Z\SA Z\SB).
   \end{eqnarray}

   The choice of $\chi$ in the initial state \eqref{eq:initial-rho} will depend on the type of interaction established between the qubits. We assume a simple interaction potential
   \begin{equation}
      V\SAB = \Omega \left(e^{i \phi\SV} \dyad{01}{10} + e^{-i \phi\SV} \dyad{10}{01}\right),
   \end{equation}
   where $\Omega$ and $\phi\SV$ are the magnitude and the phase of the coupling operator.
   $V\SAB$ becomes the standard $XY$ model with $\phi\SV=0$, and the asymmetric Dzyaloshinskii-Moriya (DM) interaction\cite{Dzyaloshinsky1958,Moriya1960} with $\phi\SV=\frac{\pi}{2}$.
   To satisfy the necessary condition for the anomalous energy exchange, $\comm{V\SAB}{\chi}\ne 0$, the correlation matrix $\chi$ must include coherence between $\ket{01}$ and $\ket{10}$. Hence, the simplest possible structure of the correlation matrix is
   \begin{equation}\label{eq:chi-matrix}
   \chi =
   \begin{pmatrix}
      \chi_{11} &           0 &          0&         0\\
      0 & \chi_{22}   & \chi_{23} &         0\\
      0 & \chi_{23}^* & \chi_{33} &         0\\
      0 &         0 &           0 &  \chi_{44}
   \end{pmatrix}.
   \end{equation}

   The correlation matrix must further satisfy three physical conditions: (a) $\rho\SAB^0$ must be positive and normalized,
   (b) $\rho\SA^0$ and $\rho\SB^0$ must remain in the Gibbs state, and  (c) $\rho\SAB^0$  must contain entanglement, which imply the mathematical conditions
   \begin{subequations}\label{eq:chi-conditions}
      \begin{equation}\label{eq:rho-positivity}
         \rho^0\SAB > 0\quad \text{and} \quad \Tr\rho^0\SAB = 1
      \end{equation}
      \begin{equation}\label{eq:traceless_chi}
         \Tr\SA \chi = \Tr\SB \chi = 0
      \end{equation}
      \begin{equation}\label{eq:concurrence-condition}
         \mathcal{C}(\rho\SAB^0) > 0,
      \end{equation}
   \end{subequations}
   where the magnitude of entanglement is measured by the concurrence $\mathcal{C}$ \cite{Hill1997}. To obtain the strongest anomalous heat current  we maximize the concurrence of the initial state under the above constraints.
   \begin{figure}
   \includegraphics[width=3.0in]{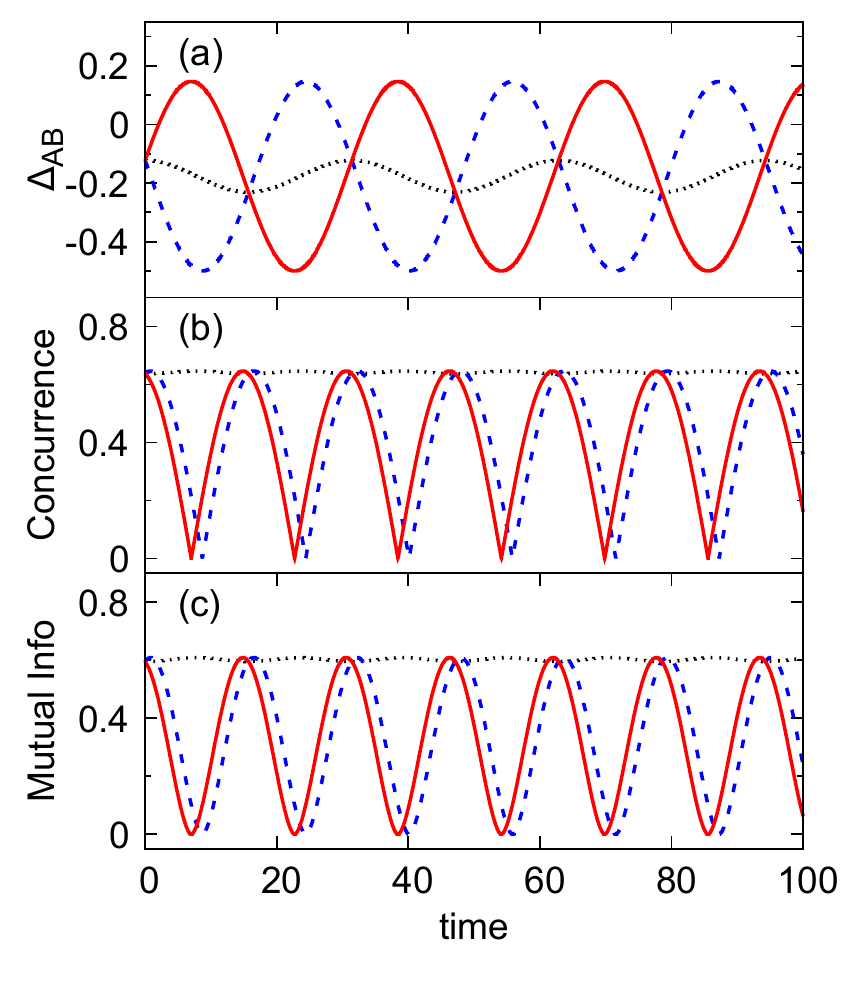}
   \caption{Evolution of (a) energy difference, (b) concurrence and  (c) mutual information of two entangled qubits at respective temperatures $T\SA=2$, $T\SB=1$ coupled through potential $V\SAB$with $\Omega=0.1$.
   The dotted black, dashed blue and solid red lines correspond to $\delta = 0,\, \frac{\pi}{2},\, -\frac{\pi}{2}$.
   For the normal energy exchange (dotted black lines), the SED never exceeds the initial value $\Delta\SAB(0)$, and the concurrence and mutual information change very little.  For the anomalous energy exchange with $\delta=-\frac{\pi}{2}$ (solid red lines), the SED
   initially increases while the concurrence and the mutual information decrease in contrast to normal energy exchange.  For the other anomalous energy exchange with $\delta=\frac{\pi}{2}$ (dashed blue lines), the SED
   initially decreases faster than that of the normal energy exchange, while the concurrence and the mutual information still decrease in almost the same way as the anomalous energy exchange with $\delta=-\frac{\pi}{2}$. In all cases, the oscillations of the SED have the same period, $\frac{\pi}{\Omega}$.
   }\label{fig:DE-phi}
   \end{figure}

   \begin{figure}
        \includegraphics[width=3.0in]{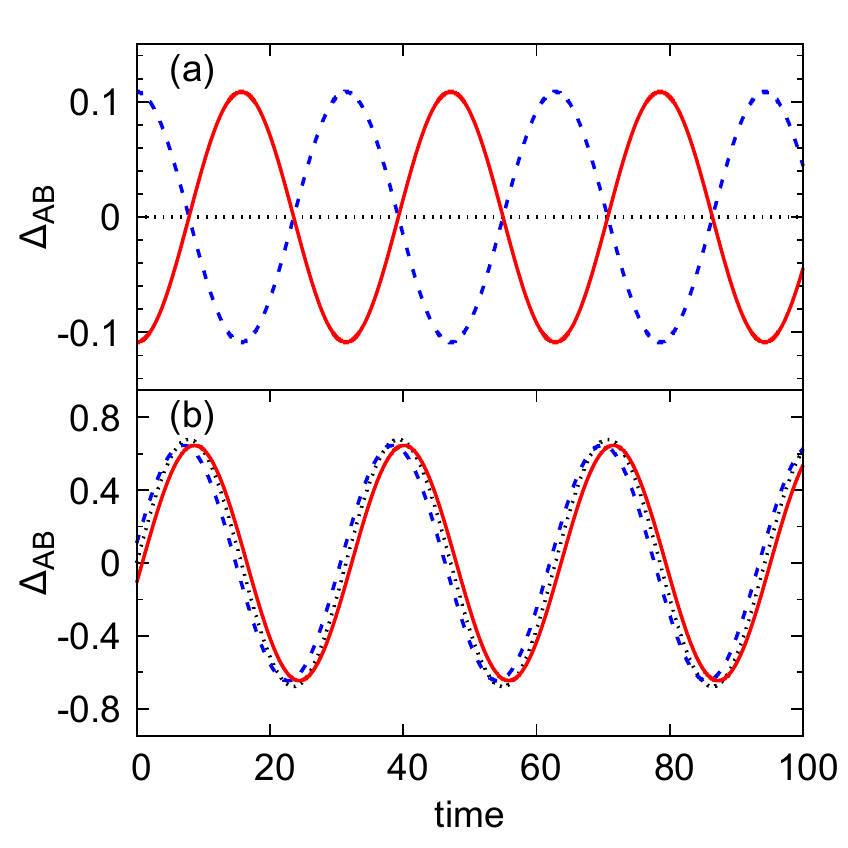}
        \caption{Local energy difference $\Delta\SAB$ is plotted for (a) normal energy exchange ($\delta=0$) and (b) anomalous energy exchange ($\delta=-\frac{\pi}{2}$) with  three different temperature gradients, \textit{dotted black line:} $T\SA=T\SB=1.5$, \textit{solid red line}: $T\SA=2.0 > T\SB=1.0$, and   \textit{dashed blue line}: $T\SA=1.0 < T\SB=2.0$.   The coupling strength, $\Omega=0.1$ is used for all cases.
        A reversal of the temperature gradient inverts the normal energy flow, but has little effect on the anomalous energy exchange.}\label{fig:DE-T}
   \end{figure}
   The optimal correlation matrix is found to be (see Appendix \ref{sec:chi})
   \begin{equation}\label{eq:optimal-chi}
      \chi =
      \begin{pmatrix}
         -\lambda_4 & 0 & 0 & 0 \\
         0 & +\lambda_4 & e^{i\phi_\chi} \sqrt{\lambda_4} & 0 \\
         0 &  e^{-i\phi_\chi} \sqrt{\lambda_4}  &+ \lambda_4 & 0 \\
         0 & 0 & 0 & -\lambda_4
      \end{pmatrix},
   \end{equation}
   where  $\phi_\chi$ is a phase between $|01\rangle$ and $|10\rangle$.  The maximum possible concurrence under the above constraints is given by
   \begin{equation}
      \mathcal{C}_0 = 2 \sqrt{\lambda_4}.
   \end{equation}
   We assume that a mapping from  $\rho\SA^0 \otimes \rho\SA^0$ to $\rho\SAB^0$ can be realized by some quantum channel. However, there may be additional restrictions. For example, if the channel is unitary, the temperature of the thermal state must be sufficiently low \cite{Huber2015}. Micadei \textit{et al.} \cite{Micadei2019} used a sequence of pulses to prepare the initial entangled state and while not necessarily optimal, were still able to measure an anomalous energy exchange.

   We solved the Liouville-von Neumann equation with the initial state \eqref{eq:product-state} and \eqref{eq:optimal-chi}, and obtained an exact solution for the time evolution of the SED between the two coupled qubits (Appendix \ref{sec:deltaERef}),
   \begin{equation}\label{eq:Delta_AB}
      \Delta\SAB(t) = \Delta\SAB(0)\sqrt{1 + \left(\frac{\mathcal{C}_0 \sin(\delta)}{\Delta\SAB(0)}\right)^2} \cos(2 \Omega t + \theta),
   \end{equation}
   where $\delta = \phi\SV-\phi_\chi$ and
   \begin{equation}
      \theta = \atan\left[\frac{\mathcal{C}_0 \sin (\delta)}{\Delta\SAB(0)}\right], \quad |\theta| < \frac{\pi}{2}.
   \end{equation}

   In the absence of initial entanglement ($\mathcal{C}_0 = 0$ and thus $\theta=0$), the SED never exceeds the initial value and normal energy exchange proceeds as expected. In the presence of an initial correlation, the energy exchange is still normal if the coupling and coherence are in phase, $\delta=0$.  Otherwise, the energy exchange is anomalous.

   Expanding Eq. \eqref{eq:DE-slope} in time the first-order term is found to be $-2\Omega\left[ \Delta\SAB^2(0)+\mathcal{C}^2_0 \sin^2 (\delta) \right]^{1/2} \sin (\theta) t$.  Hence, the SED increases only when $-\frac{\pi}{2} < \theta < 0$, which requires $\delta < 0$,  otherwise, it initially decreases faster than the normal oscillation. The sufficient condition for the anomalous energy exchange against the temperature gradient is thus $\mathcal{C}_0>0$ and $-\pi < \delta < 0$.  The fastest growth rate of the SED is obtained at $\delta=-\frac{\pi}{2}$ and it reaches the first maximum at time $\tau = \frac{|\theta|}{2\Omega}$.

   Figure \ref{fig:DE-phi} plots the SED, concurrence and mutual information in the presence of an initial correlation with $\delta=0, \pm \frac{\pi}{2}$. When the energy oscillation is normal ($\delta=0$), the concurrence and mutual information are nearly constant in time.  Otherwise ($\delta = \pm \frac{\pi}{2}$) the energy oscillation is anomalous, the entanglement and mutual information vanish as the SED reaches a maximum, indicating that the correlation is consumed to increase the SED.

    Figure \ref{fig:DE-T}, plots the normal ($\delta=0$) and anomalous ($\delta=-\frac{\pi}{2}$) energy oscillations for three different temperature gradients $T\SA=T\SB$, $T\SA > T\SB$ and $T\SA < T\SB$.  The normal energy exchange vanishes in the absence of the temperature gradient and flows in the normal direction in the presence of temperature gradient as expected. In contrast, the abnormal energy exchange is almost independent from the temperature gradient.  The amplitude of abnormal energy exchange is nearly an order of magnitude larger than that of the normal energy exchange, desirable properties for the construction of an efficient heat pump.

   \begin{figure}[t]
   \includegraphics[width=3in]{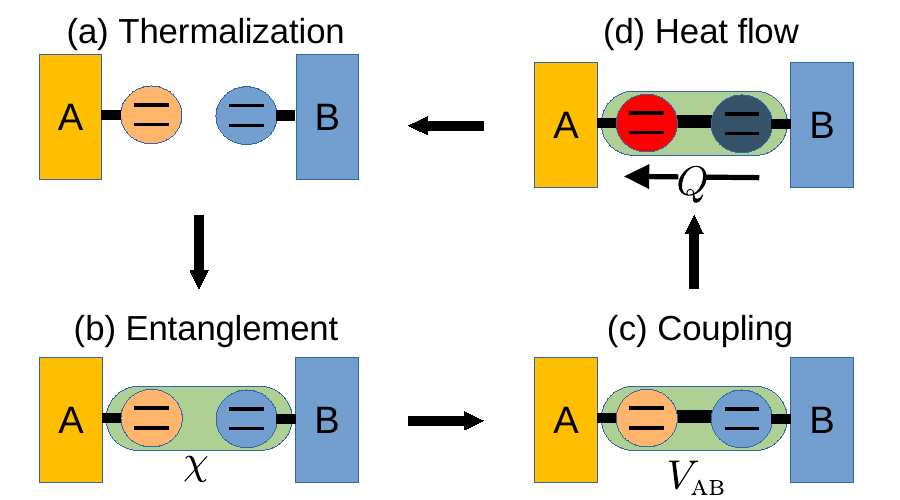}
   \caption{Heat pump cycle: (a) Qubits \texttt{A} and \texttt{B} are uncoupled from each other and thermalized with hot bath \texttt{A} and cold bath \texttt{B}, respectively. (b) Entanglement is prepared between the qubits. (c) The qubits are coupled via an interaction potential. (d) Anomalous heat flows from qubit \texttt{B} to \texttt{A} against the temperature gradient. Returning to step (a), the excess energy in qubit \texttt{A} is transferred to heat bath \texttt{A} and the energy deficiency in qubit \texttt{B} is recovered from heat bath \texttt{B} during the thermalization. }\label{fig:model}
   \end{figure}

   \section{Anomalous heat conduction between heat baths}\label{sec:heat-baths}

   The anomalous energy exchange in the isolated qubits oscillates periodically and no dissipation takes place.  In this section, we investigate the anomalous energy exchange in  a dissipative system by adding heat baths. Consider two subsystems \texttt{A} and \texttt{B} each consisting of a heat bath and a qubit as shown in Fig. \ref{fig:model}(a).  The total system \texttt{A} + \texttt{B} is isolated and thus total energy and entropy are conserved. In the absence of the qubit-qubit coupling, each qubit relaxes to a thermal state (Gibbs state with $T\SA > T\SB$) after a sufficiently long time.

   Once a local thermal equilibrium has been established in each subsystem an initial correlation $\chi$ is prepared between the qubits with $\delta=-\frac{\pi}{2}$ as discussed in Sec. \ref{sec:qubits}. As soon as the entanglement is formed, the interaction potential $V\SAB$ is turned on. Since $\tr \rho\SAB^0 V\SAB=0$, there is no energy cost to switch on the coupling, hence no work is done during this step. Energy begins flowing from qubit \texttt{B} to \texttt{A} as predicted by the previous analysis. However, the heat baths quickly destroy the coherence between the qubits and thus the anomalous energy exchange ceases at a certain time followed by normal heat conduction from \texttt{A} to \texttt{B}.  If the period of energy oscillation, $\tau_\Omega = \frac{\pi}{\Omega}$ is shorter than the decoherence time $\tau_d$, the anomalous energy exchange will reach the first maximum of its oscillation and qubit  \texttt{A} gains energy exceeding its thermal energy at $T_A$ and qubit \texttt{B} loses energy below its thermal energy at $T_B$.

   We define heat $Q\SA$ and $Q\SB$ as energy extracted from heat bath \texttt{A} and \texttt{B}, respectively \cite{Goyal2019}.  When qubit \texttt{A} acquires some excess energy due to the anomalous energy exchange, it flows into heat bath \texttt{A} and thus $Q\SA$ is negative.  Meanwhile, the deficiency of energy in qubit \texttt{B} results in positive $Q\SB$. As a whole, heat flows from bath \texttt{B} to \texttt{A} against the temperature gradient. We shall call this delivery of heat from one bath to another facilitated by the anomalous energy exchange \emph{anomalous heat}.

\begin{figure}
   \includegraphics[width=3.0in]{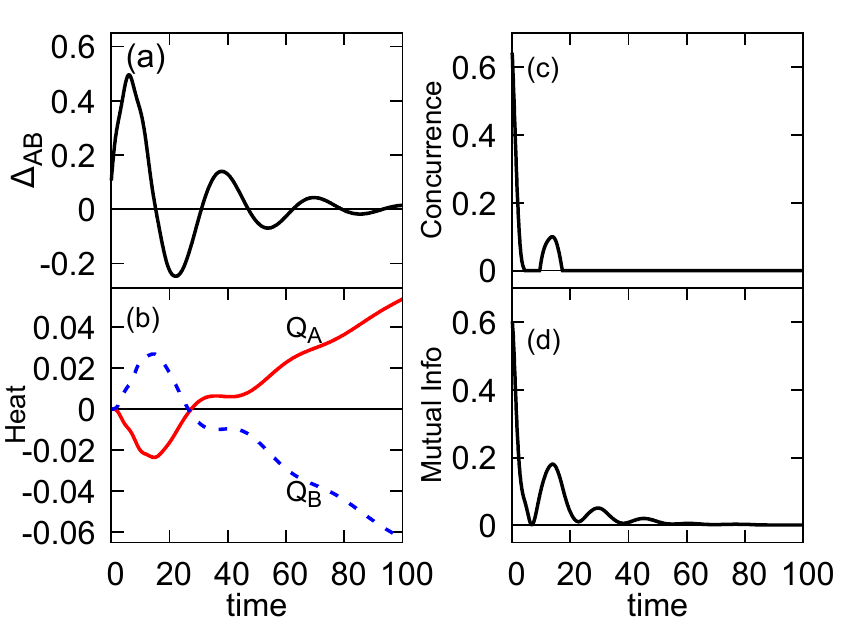}
   \caption{Transient anomalous dynamics of heat baths at respective temperatures $T\SA=2$ and $T\SB=1$ coupled through the interaction potential $V\SAB$ parametrized by $\Omega=0.1$ and $\delta=-\frac{\pi}{2}$.  Weak coupling to the heat baths ($\kappa\SA=\kappa\SB=0.01$) is assumed, where $\kappa\SA$ and $\kappa\SB$ are the coupling constants. (See Refs. \cite{Goyal2019,Kato2015}) (a) In contrast with Fig. \ref{fig:DE-T} the oscillation of the SED between qubits decays due to decoherence.  (b) Heat is initially extracted from bath \texttt{A} (\textit{solid red line}) and delivered to bath \texttt{B} (\textit{dashed blue line}) against the temperature gradient $T\SA=2$ and $T\SB=1$.  (c) Concurrence initially takes a maximum value then dies abruptly.  (d) Mutual information decays exponentially after the sudden death of the concurrence.}\label{fig:DE-HEOM}
\end{figure}

   The time evolution of the system is numerically simulated with the method of hierarchical equations of motion (HEOM) \cite{Tanimura1990}.  The detailed simulation method is described in Refs. \cite{Goyal2019,Kato2015}. In brief, the qubits are coupled to ideal Bose gases with the Drude-Lorenz spectra. The evolution of the qubits is numerically evaluated by the HEOM. Figure \ref{fig:DE-HEOM}(a) clearly shows a sharp peak in the SED near $t=\tau_\Omega$, indicating that the anomalous energy exchange is taking place even in the presence of heat baths. However, the oscillation quickly decays due to decoherence induced by the heat baths.

   Heat plotted in Fig. \ref{fig:DE-HEOM}(b) confirms that the excess energy in qubit \texttt{A} is released to heat bath \texttt{A} and at the same time qubit \texttt{B} recovers the missing energy from heat bath \texttt{B}. It also shows that the normal heat conduction takes over after one oscillation of the SED in Fig. \ref{fig:DE-HEOM}(a).

   In contrast with the case of isolated qubits discussed in the previous section, Fig. \ref{fig:DE-HEOM}(c) shows that the concurrence suddenly dies after one oscillation characteristic of the ``entanglement sudden death'' \cite{Yu2004,*Yu2006}. The mutual information plotted in figure \ref{fig:DE-HEOM}(d) survives longer than the concurrence and  gradually decays primarily due to decoherence-induced classical correlation.

   Even though this reversed heat is transient, a heat current against the temperature gradient seemingly violates the second law of thermodynamics.  If we apply the traditional theory of thermodynamics to the heat transfer between two subsystems, the entropy change in the subsystem \texttt{A(B)} is given by $\Delta S_\textsc{a(b)} + \frac{Q_\textsc{a(b)}}{T_\textsc{a(b)}}$ and the net entropy production in this interpretation is given by
   \begin{equation}\label{eq:DS1}
      \Sigma_0(t) =  \left[\Delta S\SA(t) + \frac{Q\SA}{T\SA}\right] + \left[\Delta S\SB(t) +\frac{Q\SB}{T\SB}\right],
   \end{equation}
   which can be negative in certain situations in violation of the second law.  The problem with this interpretation is that correlation between the subsystems is not taken into account. The standard theory of thermodynamics is built upon the assumption that the two subsystems are uncorrelated.   To overcome this issue, we consider the pair of qubits as a single system interacting with two heat baths simultaneously.  Then the entropy change is defined as
   \begin{equation}\label{eq:DS2}
      \Sigma(t) =  \Delta S\SAB(t) + \frac{Q\SA}{T\SA} + \frac{Q\SB}{T\SB}.
   \end{equation}
   The difference between the two definitions,  $\Sigma^0(t) - \Sigma(t) = \Delta I\SAB(t)$, is simply the mutual information.  It turns out that $\Sigma(t)$ is strictly positive if the heat baths are initially in thermal equilibrium, as shown in Appendix \ref{sec:DS}.  Hence, the definition \eqref{eq:DS2} is consistent with the second law, suggesting that the mutual information plays a significant role in thermodynamics.

\begin{figure}

   \includegraphics[width=3.0in]{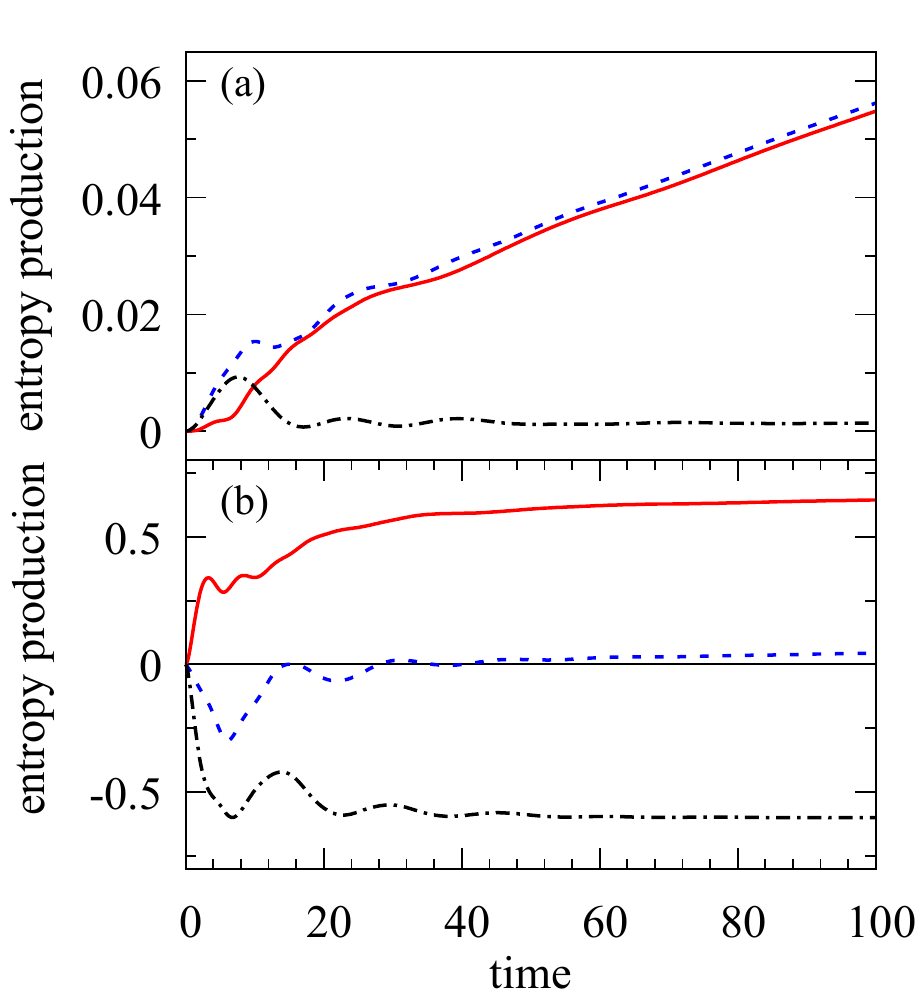}
   \caption{Two different definitions of entropy production, $\Sigma_0$ (dashed blue lines) and $\Sigma$ (solid red lines), along with the change of mutual information (dot-dashed black lines) are plotted.  For the normal heat, the two definitions of the entropy production are both positive, consistent with the second law of thermodynamics.  For the anomalous heat, $\Sigma_0$ is negative until steady normal heat is recovered, disqualifying it as the definition of entropy production.   $\Sigma$ remains positive all time, consistent with the second law. The main contribution to the entropy production is the reduction of mutual information, which is far bigger than the entropy production in the normal heat conduction.  Parameters: $\Omega=0.1$, $\delta = - \frac{\pi}{2}$ $T\SA=2$, $T\SB=1$, $\kappa\SA=\kappa\SB=0.01$. } \label{fig:DS}
   \end{figure}

   Figure \ref{fig:DS} plots numerical simulation of $\Sigma$ and $\Sigma_0$. In the absence of initial correlation [Fig. \ref{fig:DS}(a)], heat flows in the normal direction and both $\Sigma_0(t)$ and $\Sigma(t)$ remains positive at all times. However, $\Sigma_0$ is not monotonically increasing due to the transient correlation generated by the introduction of qubit-qubit coupling.  Once the steady heat is established, entropy production increases linearly.  The result is quite different when the qubits are initially correlated and the direction of heat is reversed [Fig. \ref{fig:DS}(b)].  $\Sigma_0$ is negative until the normal heat is recovered.  On the other hand, $\Sigma(t)$ remains positive and monotonically increases apart from the initial oscillation.  The entropy production with the anomalous heat is larger than that with the normal heat by an order of magnitude.  It is clear that most of the entropy production comes from the loss of the mutual information, suggesting that the dissipation associated with the anomalous heat is compensated by the loss of mutual information.

   \section{A proposed heat pump}\label{sec:heat-pump}

   Now we construct a heat pump operated cyclically by repeating the process discussed in the previous section  (see Fig. \ref{fig:model}).
   When the energy of qubit \texttt{A} reaches its first maximum at time $\tau = \frac{|\theta|}{2\Omega}$ instead of allowing normal heat to start, the qubits are disconnected by turning off the interaction potential. As a result the excess energy gained by qubit \texttt{A} is unable to return to qubit \texttt{B} and energy must flow from the qubit to heat bath \texttt{A} as heat ($Q\SA < 0$). Similarly, the energy deficiency in qubit \texttt{B} is restored by absorbing heat from heat bath \texttt{B} and thus heat $Q\SB > 0$ flows.  Eventually the subsystems reach the original thermal equilibrium. By repeating the process, heat can be cyclically extracted from the cold bath and delivered to the hot bath.

\begin{figure}
   \includegraphics[width=3.0in]{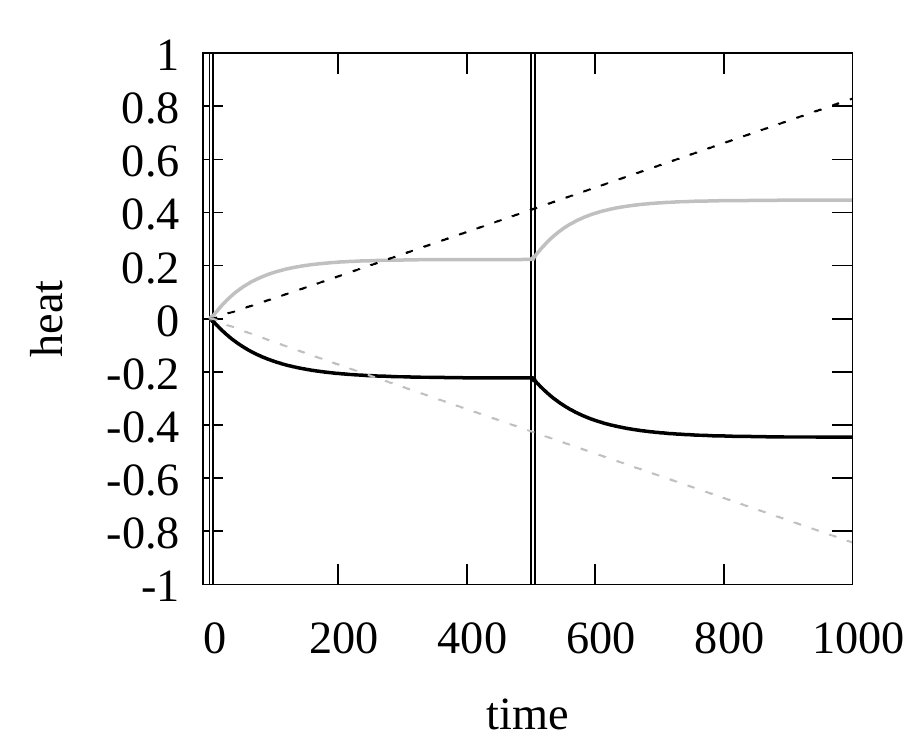}
   \caption{The performance of the heat pump obtained from numerical simulation. To eliminate the energy cost of qubit decoupling, asymmetric coupling strengths  $\kappa\SA=0.01$, and $\kappa\SB=0.023$ are used.  See Fig. \ref{fig:DS} for other parameter values. The accumulation of heat is plotted for two cycles of operation, drawn from the cold bath (solid gray line) and delivered to the hot bath (solid black line). The qubits are connected for a brief period of time $\tau = 6.1$ (narrow gaps between two vertical lines) and the system is relaxed for the period of $\tau_{\text{relax}}=500$.  See Fig. \ref{fig:DS} for other parameter values.  For comparison, the normal steady-state heat drawn from the hot bath (dashed gray line) and delivered to the cold bath (dashed black line) in the absence of initial correlation is also plotted.}\label{fig:Q-cycle}
   \end{figure}

   In Fig. \ref{fig:Q-cycle}, heat $Q\SA$ and $Q\SB$ obtained from numerical simulation are plotted for the first two cycles. The qubits are connected for a brief period of time $\tau$ (narrow shaded area) and actual heat flows from the cold to the hot bath during rethermalization. Unlike traditional heat pumps, all of the energy extracted from the cold bath is delivered to the hot bath ($\abs{Q\SA}=\abs{Q\SB}$), indicating high efficiency. For comparison, the normal heat is also plotted.  The amount of pumped heat is of the same order of the magnitude as that of the normal heat, indicating significant pumping power.

   Coupling the qubits does not require any external energy, but uncoupling the qubits requires some external work unless the interaction energy $\ev{V\SAB}$  vanishes at the moment of detachment. We note that $V\SAB$ commutes with all components of the Hamiltonian except for the interaction potential between heat baths and qubits.  Therefore, the interaction energy at the time of disconnection is negligibly small.  Furthermore it can be made infinitesimally small by adjusting the disconnection time and the coupling strength with heat baths.

   From the standard thermodynamic perspective, the efficiency of a heat pump is measured by the coefficient of performance (COP)
   \begin{equation}\label{eq:COP0}
      \eta_\textsc{cop} = \frac{Q\SB}{W} = \frac{Q\SB}{Q\SA+Q\SB},
   \end{equation}
   which diverges when evaluated for our heat pump because no work is done on the system, and thus $Q\SA+Q\SB=0$ at the end of every cycle. The second law of thermodynamics demands that the efficiency should be bounded by the Carnot efficiency $\eta_\textsc{cop}^\text{max} = \frac{T\SB}{T\SA-T\SB}$.  It has been shown that quantum coherency or correlation cannot violate the Carnot statement of the second law \cite{Gardas2015}. To reestablish the second law, we must change the definition of COP. Recalling the definition of entropy production \eqref{eq:DS2} and further noting that $\Delta S\SA=\Delta S\SB=0$ at the end of one cycle, the second law becomes
   \begin{equation}\label{eq:DS-cycle}
      I_0 - \frac{\bar{Q}\SA}{T\SA} - \frac{\bar{Q}\SB}{T\SB}  \ge 0,
   \end{equation}
   where $I_0$ is the initial mutual information and $\bar{Q}$ is heat per cycle.  Rearranging Eq. \eqref{eq:DS-cycle} yields,
   \begin{equation}\label{eq:COP1}
      \frac{T\SB}{T\SA-T\SB} \ge \frac{\bar{Q}\SB}{T\SA I_0}.
   \end{equation}
   The direct comparison of \eqref{eq:COP1} with the standard Carnot efficiency suggests that an effective work may be defined as $W = T\SA I_0$ \cite{Bera2017}.  Substituting the results of the simulation, we find $\frac{\bar{Q}\SB}{T\SA I_0} = 0.186$ which is lower than the Carnot efficiency $\eta_\textsc{cop}^\text{max}=1$, satisfying the inequality \eqref{eq:COP1}.

   \section{Discussion and Conclusions}
   We first investigated the anomalous energy exchange between an entangled pair of qubits locally in thermal states with different temperature.  We have shown that the direction and magnitude of the energy flow can be controlled by tuning the phases in the initial coherence and coupling operator as well as the concurrence of the initial state. The direction of the anomalous energy flow is nearly independent of the temperature gradient

   Second, we investigated the effect of decoherence on the anomalous energy exchange by coupling independent thermal baths to the qubits.  The anomalous energy exchange quickly disappears as the entanglement vanishes due to decoherence.  However, as the qubits are thermalized, the excess energy in the qubits dissipates to the baths as heat, reversing the direction of heat against the temperature gradient.

   Based on the anomalous heat, we have constructed a heat pump.  By tuning the parameters, namely, the concurrence of the initial state, the phases and the coupling strength, we have shown numerically that rather strong heat current is created against the temperature gradient. If we assume that the initial mutual information multiplied by temperature is equivalent to external work, the coefficient of performance obtained from simulation is consistent with the second law of thermodynamics.

   To operate the heat pump cyclically, the initial correlation must be periodically reestablished. Micadei \textit{et al.} \cite{Micadei2019} used a sequence of single-qubit rotations and two-qubit interactions to prepare the fuel state experimentally and were able to realize some correlation sufficient to reverse the direction of the energy exchange between qubits. Other possibilities include the entanglement-preserving local thermalization channel \cite{Hsieh2020} or entanglement swapping in the presence of thermal environments \cite{Sen2005,Dajka2009,Miao2013,Zou2015,Kirby2016,Nourmandipour2016}.

   It has been shown that coherence can be created or increased using the temperature gradient (heat) as resources \cite{Mukhopadhyay2018,Manzano2019}. It would be interesting to investigate the time reversed process of our proposed heat pump. If the heat from a hot to a cold bath is increased compared with normal heat conduction, is it possible to amplify coherence?

   \begin{acknowledgments}
   We would like to thank Garrett Van Dyke for executing the simulation and contributing to the discussion.
   \end{acknowledgments}

   \appendix

   \section{Optimal Correlation Matrix}\label{sec:chi}

   In this appendix, we determine the optimal correlation matrix \eqref{eq:optimal-chi} using the conditions \eqref{eq:chi-conditions}.
   Since the matrix \eqref{eq:chi-matrix} takes a so-called X form \cite{HashemiRafsanjani2012}, we can easily evaluate the conditions.
   The positivity condition \eqref{eq:rho-positivity} is satisfied by
   \begin{subequations}\label{eq:cond1}
   \begin{equation}
       \lambda_i + \chi_{ii} \ge 0, \quad \forall i
   \end{equation}
   \begin{equation}
       \sqrt{(\lambda_2+\chi_{22})(\lambda_3+\chi_{33})} \ge \abs{\chi_{23}}.
   \end{equation}
   \end{subequations}
    The traceless conditions \eqref{eq:traceless_chi} constrains the diagonal elements by $\chi_{11}=-\chi_{22}=-\chi_{33}=\chi_{44}$, which reduces the conditions \eqref{eq:cond1} to
    \begin{subequations}\label{eq:cond2}
       \begin{equation}\label{eq:cond2-1}
          \min\left(\lambda_2,\lambda_3\right) \ge \chi_{11} \ge - \lambda_4
       \end{equation}
       and
       \begin{equation}\label{eq:cond2-2}
          \sqrt{(\lambda_2 - \chi_{11})(\lambda_3 - \chi_{11})} \ge \abs{\chi_{23}}.
       \end{equation}
    \end{subequations}
    Now, we have only two parameters to determine, $\chi_{11}$ and $\chi_{23}$.

\begin{figure}
       \includegraphics[width=2.5in]{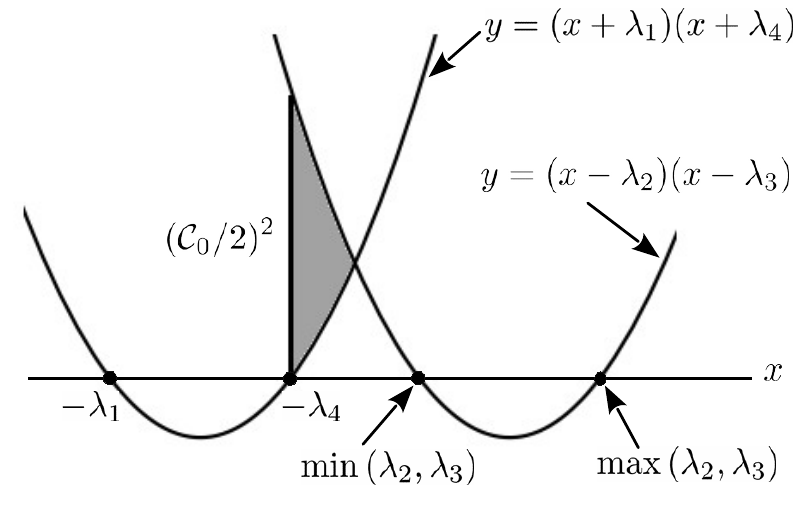}
       \caption{The schematic representation of three conditions \eqref{eq:cond2-1}, \eqref{eq:cond2-2}, and \eqref{eq:cond3}.  $x$ and $y$ represent $\chi_{11}$ and $|\chi_{23}|^2$, respectively.  The shaded region satisfies all conditions. The vertical edge of the region corresponds to the square of maximum concurrence.}\label{fig:chi-conditions}
    \end{figure}

    Concurrence is found to be
    \begin{equation}
       \mathcal{C}(\rho) = 2 \max \{0, \abs{\chi_{23}} - \sqrt{(\lambda_1+\chi_{11})(\lambda_4+\chi_{11})}\}
    \end{equation}
    and thus entanglement is formed only if
    \begin{equation}\label{eq:cond3}
       \abs{\chi_{23}} > \sqrt{(\lambda_1+\chi_{11})(\lambda_4+\chi_{11})}.
    \end{equation}
   Combining the conditions  \eqref{eq:cond2} and \eqref{eq:cond3}, the parameters take values in the shaded area in Fig. \ref{fig:chi-conditions}.

   Finally, we obtain the maximum concurrence $\mathcal{C}_0 = 2 \sqrt{\lambda_4}$ with $\chi_{11}=-\lambda_4$ and $\abs{\chi_{23}} = \sqrt{\lambda_4}$  (see Fig. \ref{fig:chi-conditions}). Now, we have the optimal correlation matrix \eqref{eq:optimal-chi}.

   \section{Unitary Evolution of Coupled Qubits}\label{sec:deltaERef}

   We solve the Liouville-von Neumann equation $i \dot{\rho} =  \comm{( H\SA + H\SB + V\SAB)}{\rho}$ where the Hamiltonians are defined in Section \ref{sec:qubits}. The initial state \eqref{eq:initial-rho} is specifically written in the matrix forms \eqref{eq:product-state} and \eqref{eq:optimal-chi}.  Noting that $\rho_{11}$ and $\rho_{44}$ are independent from other matrix elements, they are constant in time and thus $\rho_{11}(t) = \lambda_1-\lambda_2$ and $\rho_{44}(t)=0$. The remaining equations of motion are
   \begin{subequations}
      \begin{equation}
         i \dot{\rho}_{22}= \Omega \left[ e^{i\phi\SV}\rho_{32} - e^{-i\phi\SV}\rho_{23}\right],
      \end{equation}
      \begin{equation}
         i \dot{\rho}_{23} = \Omega e^{i \phi\SV} \left(\rho_{33}-\rho_{22}\right)
      \end{equation}
   \end{subequations}
   with the conservation of two quantities: $\dot{\rho}_{22}+ \dot{\rho}_{33} =0$ and $ e^{i \phi\SV} \dot{\rho}_{32} + e^{-i\phi\SV} \dot{\rho}^*_{23}=0$.
   We obtain the following solution:
   \begin{subequations}
   \begin{equation}
      \begin{split}
      \rho_{22}(t) & = \frac{1}{2} \left[ \lambda_2 + \lambda_3 + 2 \lambda_4 \right . \\
       & \left .- \Delta\SAB(0) \cos(2\Omega t) + \sin(\delta) \mathcal{C}_0 \sin(2\Omega t)\right],
      \end{split}
   \end{equation}
   \begin{equation}
      \begin{split}
      \rho_{33}(t) & = \frac{1}{2} \left[ \lambda_2 + \lambda_3 + 2 \lambda_4 \right . \\
      & \left .+ \Delta\SAB(0) \cos(2\Omega t) - \sin(\delta) \mathcal{C}_0 \sin(2\Omega t)\right],
      \end{split}
   \end{equation}
   \begin{equation}
      \begin{split}
         \rho_{23}(t) &= \rho^*_{32}(t) = \frac{e^{i\phi\SV}}{2} \left[ \mathcal{C}_0\cos(\delta)  \right . \\
         & \left .-i \left\{\sin(\delta)\mathcal{C}_0 \cos(2\Omega t) - \Delta\SAB(0) \sin(2\Omega t)\right\}\right].
      \end{split}
   \end{equation}
   \end{subequations}

   Next we evaluate the energy expectation values $E\SA = \Tr \left[ (H\SA \otimes I) \rho \right]$ and $E\SB = \Tr \left[ (I \otimes H\SB) \rho\right]$.  Using the above solutions,  they are evaluated as
   \begin{equation}
      \begin{split}
          E_{\textsc{a}/\textsc{b}}(t) & = \frac{1}{2}\left[\lambda_1-\lambda_4 \right .\\
          & \left. \mp \Delta\SAB(0) \cos(2 \Omega t) \mp \sin(\delta) \mathcal{C}_0 \sin(2\Omega t)\right],
      \end{split}
   \end{equation}
    from which we obtain Eq. \eqref{eq:Delta_AB}.

   \section{Entropy production}\label{sec:DS}

   We show that Eq. \eqref{eq:DS2} is non-negative.   Consider a system \texttt{s} in contact with hot bath \texttt{h} and cold bath \texttt{c}.  The Hilbert space of the total system is given by $\mathcal{H}_\text{s} \otimes \mathcal{H}_\text{h} \otimes \mathcal{H}_\text{c}$. We assume that they are initially uncorrelated and the heat baths are in thermal equilibrium,  thus the initial density can be written as
   \begin{equation}
   \rho(0) = \rho_\text{s}(0) \otimes \rho_\text{h}^\textsc{g} \otimes \rho_\text{c}^\textsc{g},
   \end{equation}
   where $\rho_\text{s}(0)$ is an arbitrary initial system density  and the initial density of bath $k=\text{h, c}$ takes the Gibbs form $\rho_k^\textsc{g} = e^{-\beta_k H_\ell} / Z_k$ with the bath Hamiltonian $H_k$, inverse temperature $\beta_k$, and partition function $Z_k$.

   We define heat as energy leaving the heat bath:

   \begin{eqnarray}
       Q_k(t) &=&  \Tr_k  \left( \rho_k^\textsc{g} H_k  \right) - \Tr_k  \left( \rho_k(t) H_k \right)\nonumber\\
       &=& -\frac{1}{\beta_k}  \left [
         \Tr_k \left( \rho_\ell^\textsc{g} \ln \rho_\ell^\textsc{g} \right)
         -\Tr_k \left( \rho_k (t) \ln \rho_k^\textsc{g} \right)
          \right].\label{eq:heat}
   \end{eqnarray}
The change in the entropy of the system is defined with the von Neumann entropy:
\begin{equation}
    \Delta S_\text{s} = -\Tr_\text{s} \left(\rho_\text{s}(t) \ln \rho_\text{s}(t)\right) + \Tr_\text{s} \left(\rho_\text{s}(0) \ln \rho_\text{s}(0)\right).\label{eq:DeltaSS}
\end{equation}

Rewriting Eqs. \eqref{eq:heat} and \eqref{eq:DeltaSS} using trace over the whole Hilbert space, substituting them to the definition of entropy production \eqref{eq:DS2}, rearranging terms, and using the conservation of total entropy,  we obtain

\begin{align}
    \Sigma &= \Delta S_\text{s} - \beta_\text{h} Q_\text{h} - \beta_\text{c} Q_\text{c} \nonumber\\
    & = - \Tr \left[\rho(t) \ln \left(\rho_\text{s}(t) \otimes \rho_\text{h}^\textsc{g} \otimes \rho_\text{c}^\textsc{g} \right) \right] \nonumber \\
    & + \Tr \left[ \left(\rho_\text{s}(0) \otimes \rho_\text{h}^\textsc{g} \otimes \rho_\text{c}^\textsc{g} \right) \ln \left(\rho_\text{s}(0) \otimes \rho_\text{h}^\textsc{g} \otimes \rho_\text{c}^\textsc{g}\right) \right] \nonumber \\
    & = S\left(\rho(t) \| \rho_\text{s}(t) \otimes \rho_\text{h}^\textsc{g} \otimes \rho_\text{c}^\textsc{g} \right) \ge 0
\end{align}
which shows that Eq. \eqref{eq:DS2} is always non-negative for $t \ge 0$.

    \bibliographystyle{apsrev4-2}
    \bibliography{references}

\end{document}